\documentclass[aps,showpacs]{revtex4}
\usepackage{graphicx}
\begin{document}

\title{The transverse energy and the charged particle multiplicity
in a statistical model with expansion}
\thanks{Presented at the
XLIV Cracow School of Theoretical Physics, Zakopane, Poland, May
28-June 6, 2004}
\author{Dariusz Prorok}
\affiliation{Institute of Theoretical Physics, University of
Wroc{\l}aw,\\ Pl.Maksa Borna 9, 50-204  Wroc{\l}aw, Poland}
\date{October 1, 2004}

\begin{abstract}
Global variables such as the transverse energy and the charged
particle multiplicity and their ratio are evaluated, in a
statistical model with expansion, for heavy-ion collisions from
AGS to RHIC at $\sqrt{s_{NN}}=200$ GeV. Full description of decays
of hadron resonances is applied. The predictions of the model done
at the freeze-out parameters established independently from
observed particle yields and $p_{T}$ spectra agree well with the
experimental data. However, some (explicable) overestimation of
the ratio has been found.
\end{abstract}

 \maketitle

\section {Introduction }

The statistical model has been applied successfully in description
of the soft part of particle production in heavy-ion collisions
\cite{Braun-Munzinger:1994xr,Braun-Munzinger:1995bp,Cleymans:1996cd,Stachel:wh,Braun-Munzinger:1999qy,Becattini:2000jw,Braun-Munzinger:2001ip,Florkowski:2001fp,Broniowski:2001we,Broniowski:2001uk,Baran:2003nm,Broniowski:2002nf,Michalec:2001um,Broniowski:2002am}.
Namely, the particle yield ratios and $p_{T}$ spectra were fitted
accurately with the use of only four parameters (for particle
ratio fits two parameters are enough: the temperature $T$ and the
baryon number chemical potential $\mu_{B}$). Now, those parameters
will be used to evaluate global observables: the transverse energy
density $dE_{T}/d\eta$, the charged particle multiplicity density
$dN_{ch}/d\eta$ and their ratio (for details see
\cite{Prorok:2004af}). The advantage of such an approach is based
on the fact that the transverse energy and charged particle
multiplicity measurements are independent of hadron spectroscopy
(in particular, no particle identification is necessary),
therefore they could be used as an additional test of the
self-consistency of a statistical model.

The experimentally measured transverse energy is defined as

\begin{equation}
E_{T} = \sum_{i = 1}^{L} \hat{E}_{i} \cdot \sin{\theta_{i}} \;,
\label{Etdef}
\end{equation}

\noindent where $\theta_{i}$ is the polar angle, $\hat{E}_{i}$
denotes $E_{i}-m_{N}$ ($m_{N}$ means the nucleon mass) for baryons
and the total energy $E_{i}$ for all other particles, and the sum
is taken over all $L$ emitted particles \cite{Adcox:2001ry}.
Additionally, in the case of RHIC at $\sqrt{s_{NN}}=200$ GeV,
$E_{i}+m_{N}$ is taken instead of $E_{i}$ for antibaryons
\cite{Bazilevsky:2002fz}.

The statistical model with single freeze-out is used (for details
see \cite{Broniowski:2002nf}). The model very well reproduces
ratios and $p_{T}$ spectra of particles measured at RHIC
\cite{Florkowski:2001fp,Broniowski:2001we,Broniowski:2001uk}. The
main assumption of the model is the simultaneous occurrence of
chemical and thermal freeze-outs, which means that the possible
elastic interactions after the chemical freeze-out are neglected.
The conditions for the freeze-out are expressed by values of two
independent thermal parameters: $T$ and $\mu_{B}$. The strangeness
chemical potential $\mu_{S}$ is determined from the requirement
that the overall strangeness of the gas equals zero.

The actually detected (stable) particles have two sources:
(\textit{a}) a thermal gas and (\textit{b}) secondaries produced
by decays and sequential decays of primordial resonances. The
distributions of particles from source (\textit{a}) are given by a
Bose-Einstein or a Fermi-Dirac distribution at the freeze-out
point. The distributions of secondaries (source (\textit{b})) can
be obtained from the elementary kinematics of a many-body decay or
from the superposition of two or more such decays (for details see
\cite{Broniowski:2002nf,Prorok:2004af}). In the following, all
possible (2-, 3- and 4-body) decays with branching ratios not less
than $1 \%$ are considered. Also almost all cascades (with the
exclusion of a very few like 4- or 5-step ones such that they
proceed via at least one 3- or 4-body decay) are taken into
account. It should be stressed that all contributions from weak
decays are included in the presented analysis.

\section { The foundations of the single-freeze-out model }
\label{Foundat}

The foundations of the model are as follows: (\textit{a}) the
chemical and thermal freeze-outs take place simultaneously,
(\textit{b}) all confirmed resonances up to a mass of $2$ GeV from
the Particle Data Tables \cite{Hagiwara:fs} are taken into
account, (\textit{c}) a freeze-out hypersurface is defined by the
equation

\begin{equation}
\tau = \sqrt{t^{2}-r_{x}^{2}-r_{y}^{2}-r_{z}^{2}}= const \;,
\label{Hypsur}
\end{equation}

\noindent (\textit{d}) the four-velocity of an element of the
freeze-out hypersurface is proportional to its coordinate

\begin{equation}
u^{\mu}={ {x^{\mu}} \over \tau}= {t \over \tau}\; \left(1,{
{r_{x}} \over t},{{r_{y}} \over t},{{r_{z}} \over t}\right) \;,
\label{Velochyp}
\end{equation}

\noindent (\textit{e}) the transverse size is restricted by the
condition $r=\sqrt{r_{x}^{2}+r_{y}^{2}}< \rho_{max}$. In this way
one has two additional parameters of the model, $\tau$ and
$\rho_{max}$, connected with the geometry of the freeze-out
hypersurface.

The maximum transverse-flow parameter (or the surface velocity) is
given by

\begin{equation}
\beta_{\perp}^{max}= { {\rho_{max}/\tau} \over
{\sqrt{1+(\rho_{max}/\tau)^{2}}}}\;. \label{Betmax}
\end{equation}

The invariant distribution of the measured particles of species
$i$ has the form \cite{Broniowski:2001we,Broniowski:2001uk}

\begin{equation}
{ {dN_{i}} \over {d^{2}p_{T}\;dy} }=\int
p^{\mu}d\sigma_{\mu}\;f_{i}(p \cdot u) \;, \label{Cooper}
\end{equation}

\noindent where $d\sigma_{\mu}$ is the normal vector on a
freeze-out hypersurface, $p \cdot u = p^{\mu}u_{\mu}$ , $u_{\mu}$
is the four-velocity of a fluid element and $f_{i}$ is the final
momentum distribution of the particle in question. The final
distribution means here that $f_{i}$ is the sum of primordial and
simple and sequential decay contributions to the particle
distribution (for details see \cite{Broniowski:2002nf}).

The pseudorapidity density of particle species $i$ is given by

\begin{equation}
{ {dN_{i}} \over {d\eta} } = \int d^{2}p_{T}\; {{dy} \over {d\eta}
} \; { {dN_{i}} \over {d^{2}p_{T}\;dy} }= \int d^{2}p_{T}\; {p
\over {E_{i}} } \; { {dN_{i}} \over {d^{2}p_{T}\;dy} }\;.
\label{Partdens}
\end{equation}

\noindent Analogously, the transverse energy pseudorapidity
density for the same species can be written as

\begin{equation}
{ {dE_{T,i}} \over {d\eta} } = \int d^{2}p_{T}\; \hat{E}_{i} \cdot
{{p_{T}} \over p} \; {{dy} \over {d\eta} }\; { {dN_{i}} \over
{d^{2}p_{T}\;dy} }= \int d^{2}p_{T}\;{p_{T}} \; { {\hat{E}_{i}}
\over {E_{i}} }\; { {dN_{i}} \over {d^{2}p_{T}\;dy} }\;.
\label{Etraden}
\end{equation}

\noindent For the quantities at midrapidity one has

\begin{equation}
{ {dN_{i}} \over {d\eta} }\;\Big\vert_{mid}= \int d^{2}p_{T}\;{
{dN_{i}} \over {d^{2}p_{T}\;dy} }\; {
{\sqrt{p_{T}^{2}+v_{c.m.s}^{2}m_{i}^{2}}} \over {m_{T}} } \;,
\label{Partdenmid}
\end{equation}

\begin{equation}
{ {dE_{T,i}} \over {d\eta} }\;\Big\vert_{mid} = \cases{ \int
d^{2}p_{T}\;{p_{T}} \;{ {dN_{i}} \over {d^{2}p_{T}\;dy} }\;  {
{m_{T}-\sqrt{1-v_{c.m.s}^{2}}m_{N}} \over {m_{T}} }, i=nucleon
 \cr \cr \int d^{2}p_{T}\;{p_{T}} \;{ {dN_{i}} \over
{d^{2}p_{T}\;dy} }, i \neq nucleon\;.} \label{Etdenmid}
\end{equation}

\noindent where $v_{c.m.s}$ is the velocity of the center of mass
of two colliding nuclei with respect to the laboratory frame (only
for RHIC $v_{c.m.s}=0$). Note that for RHIC at $\sqrt{s_{NN}}=200$
GeV there is the third possibility in Eq.~(\ref{Etdenmid}): if
$i=antinucleon$, the analogous formula as for $i=nucleon$ but with
$m_{T}+m_{N}$ in the numerator holds.

The overall charged particle and transverse energy densities can
be expressed as

\begin{equation}
{ {dN_{ch}} \over {d\eta} }\;\Big\vert_{mid}= \sum_{i \in B} {
{dN_{i}} \over {d\eta} }\;\Big\vert_{mid}\;, \label{Nchall}
\end{equation}

\begin{equation}
{ {dE_{T}} \over {d\eta} }\;\Big\vert_{mid}= \sum_{i \in A} {
{dE_{T,i}} \over {d\eta} }\;\Big\vert_{mid} \;, \label{Etall}
\end{equation}

\noindent where $A$ and $B$ ($B \subset A$) denote sets of species
of finally detected particles, namely the set of charged particles
$B=\{\pi^{+},\; \pi^{-},\; K^{+},\; K^{-},\; p,\; \bar{p}\}$,
whereas $A$ also includes photons, $K_{L}^{0},\; n$ and
$\bar{n}\;$ \cite{Adcox:2001ry}.

\section {Results}
\label{Finl}

The general scheme reviewed in the previous section was formulated
originally for RHIC
\cite{Florkowski:2001fp,Broniowski:2001we,Broniowski:2001uk} and
then applied for SPS \cite{Broniowski:2002am}. Here, this method
will be used also for the AGS case. But the different model of the
freeze-out hypersurface was applied for the description of $p_{T}$
spectra there \cite{Braun-Munzinger:1994xr,Stachel:wh}. In that
model (for details see \cite{Schnedermann:1993ws}), the freeze-out
happens instantaneously in the $r$ direction, \emph{i.e.} at a
constant value of $t$ (\emph{not} at a constant value of $\tau$ as
here). The parameters connected with the expansion are the surface
velocity $\beta_{\perp}^{max}$ and $\rho_{max}$. Therefore, the
implementation of values of $\beta_{\perp}^{max}$ obtained within
that model into the presented one is entirely \emph{ad hoc},
nevertheless it works surprisingly well. To put values of
$\beta_{\perp}^{max}$ from
\cite{Braun-Munzinger:1994xr,Stachel:wh} into formulae of
sect.~\ref{Foundat}, one should invert Eq.~(\ref{Betmax}) to
obtain

\begin{equation}
{{\rho_{max}} \over {\tau} }=  { {\beta_{\perp}^{max}} \over
{\sqrt{1-(\beta_{\perp}^{max})^{2}}}}\;. \label{Rhmaxta}
\end{equation}

\noindent It should be recalled here that the value of $\tau$
itself is not necessary to calculate the transverse energy per
charged particle, since this parameter cancels in the ratio.

The final results of numerical estimates of
$dE_{T}/d\eta\vert_{mid}$ and $dN_{ch}/d\eta\vert_{mid}$ together
with the corresponding experimental data are listed in
Table\,\ref{Table1}. To make predictions for the AGS case it has
been assumed that the maximal transverse size $\rho_{max}$ equals
the average of radii of two colliding nuclei and the nucleus
radius has been expressed as $R_{A}=r_{0}A^{{1 \over
3}},\;r_{0}=1.12$ fm. Generally, the estimates agree well with the
data. However, for RHIC the $11\%-16\%$ underestimation of the
charged particle density has been found. But this simply reflects
the existing inconsistency in measurements of the charged particle
multiplicity at RHIC. Namely, the sum of integrated charged hadron
yields \cite{Adcox:2001mf}, after converting to $dN_{ch}/d\eta$
\cite{Bazilevsky:2002fz}, is substantially less then the directly
measured $dN_{ch}/d\eta\vert_{mid}$ \cite{Adcox:2000sp}. This is
shown explicitly in the last column of Table\,\ref{Table1}. But
values of the sum agree very well with the model predictions.
Since the geometric parameters were established from the fits to
the same $p_{T}$ spectra, the agreement had to be obtained. Also
for AGS the results agree qualitatively well with the data, in
spite of the roughness of the method applied for this case. The
overall error of evaluations of transverse energy and charged
particle densities is about $0.5\%$ and is caused by: (\textit{a})
omission of the most complex cascades; (\textit{b})
simplifications in numerical procedures for more involved
cascades. The velocity of the center of mass of two colliding
nuclei, $v_{c.m.s}$, equals: 0 for RHIC, 0.994 for SPS Pb-Pb
collisions at $158 \cdot A$ GeV, 0.918 for AGS Au-Au collisions at
$11 \cdot A$ GeV and 0.678 for AGS Si-Pb collisions at $14.6 \cdot
A$ GeV.
\begin{table}
\caption{\label{Table1} Values of $dE_{T}/d\eta\vert_{mid}$ and
$dN_{ch}/d\eta\vert_{mid}$ calculated in the framework of the
statistical model with expansion. In the first column thermal and
geometric parameters are listed for the corresponding collisions.
In the third and last column experimental data for the most
central collisions are given.}
\begin{ruledtabular}
\begin{tabular}{c c c c c c} \hline Collision case &
\multicolumn{2}{c}{$dE_{T}/d\eta\vert_{mid}$
[GeV]} & & \multicolumn{2}{c}{$dN_{ch}/d\eta\vert_{mid}$}
\\
\cline{2-3}\cline{5-6} & Theory & Experiment & & Theory &
Experiment
\\
\hline Au-Au at RHIC at $\sqrt{s_{NN}}=200$ GeV: & & & & &
\\
 $T
= 165.6$ MeV, $\mu_{B} = 28.5$ MeV & 585 \footnote{For the
modified definition of $E_{T}$, \emph{i.e.} $E_{i}+m_{N}$ is taken
instead of $E_{i}$ for antibaryons, see
Eq.~(\protect\ref{Etdef}).} & $597 \pm 34$
\protect\cite{Bazilevsky:2002fz} & & 589 & $699 \pm 46$
\protect\cite{Bazilevsky:2002fz}
\\
 $\rho_{max} = 7.15$ fm, $\tau
= 7.86$ fm ($\beta_{\perp}^{max} = 0.67$)
\protect\cite{Baran:2003nm} & & & & & $579 \pm 29$ \footnote{For
the charged particle multiplicity expressed as the sum of
integrated charged hadron yields.}
\\
 & & & & & \protect\cite{Adler:2003cb} \\  Au-Au at
RHIC at $\sqrt{s_{NN}}=130$ GeV: & & & & &
\\
 $T = 165$ MeV,
$\mu_{B} = 41$ MeV & 507 & $503 \pm 25$
\protect\cite{Adcox:2001ry} & & 555 & $622 \pm 41$
\protect\cite{Adcox:2000sp}
\\
 $\rho_{max} = 6.9$ fm, $\tau =
8.2$ fm ($\beta_{\perp}^{max} = 0.64$)
\protect\cite{Broniowski:2002nf} & & & & & $568 \pm 47\;^{b}$
\\
 & & & & & \protect\cite{Adcox:2001mf}
\\
 Pb-Pb
at SPS: & & & & & \\
 $T = 164$ MeV, $\mu_{B} = 234$ MeV & 447 &
$363 \pm 91$ \protect\cite{Aggarwal:2000bc} & & 476 &
$464_{-13}^{+20}$ \protect\cite{Aggarwal:2000bc}
\\ $\rho_{max} =
6.45$ fm, $\tau = 5.74$ fm ($\beta_{\perp}^{max} = 0.75$)
\protect\cite{Michalec:2001um,Broniowski:2002am} & & & & & \\  & &
& & & \\ Au-Au at AGS: & & & & & \\ $T = 130$ MeV, $\mu_{B} =
540$ MeV & 224 & $\approx 200$ \protect\cite{Barrette:pm} & & 271
& $\approx 270$ \protect\cite{Barrette:1994kr} \\
$\beta_{\perp}^{max} = 0.675$, $\rho_{max} = 6.52$ fm
\protect\cite{Braun-Munzinger:1994xr,Stachel:wh} & & & & & \\ & &
& & & \\  Si-Pb at AGS: & & & & & \\  $T = 120$ MeV, $\mu_{B} =
540$ MeV & 57 & $\approx 62$ \protect\cite{Barrette:1994kr} & & 91
& $\approx 115-120$ \\ $\beta_{\perp}^{max} = 0.54$, $\rho_{max} =
5.02$ fm \protect\cite{Braun-Munzinger:1994xr,Stachel:wh} & & & & &
\protect\cite{Barrette:1994kr} \\
\hline
\end{tabular}
\end{ruledtabular}
\end{table}

Values of the ratio $dE_{T}/d\eta\vert_{mid} /
dN_{ch}/d\eta\vert_{mid}$ can be also calculated. They are
collected in Table\,\ref{Table2}, together with the corresponding
data. The overall overestimation of the order of $15\%$ has been
obtained. In the RHIC case this is the result of the
underestimation of $dN_{ch}/d\eta\vert_{mid}$, which has been
explained earlier. But when in the denominator of the experimental
ratio, $dN_{ch}/d\eta\vert_{mid}$ from the summing up of
integrated hadron yields is put, the theoretical predictions agree
very well with the data. Note that the similar inconsistency in
charged particle measurements could have also been the origin of
the discrepancy between model and experimental values of
$dN_{ch}/d\eta\vert_{mid}$ seen in the AGS Si-Pb case. For SPS,
the result agrees with the experimental value within errors. The
overall error of model evaluations of the ratio is less than $1
\%$.
\begin{table}
\caption{\label{Table2} Values of the ratio
$dE_{T}/d\eta\vert_{mid}/ dN_{ch}/d\eta\vert_{mid}$ calculated in
the framework of the statistical model with expansion. In the last
column experimental data for the most central collisions are
given.}
\begin{ruledtabular}
\begin{tabular}{c c c} \hline { }
& \multicolumn{2}{c}{} \\ Collision case & \multicolumn{2}{c}{
$dE_{T}/d\eta\vert_{mid}/ dN_{ch}/d\eta\vert_{mid}$ [GeV]} \\  &
\multicolumn{2}{c}{} \\ \cline{2-3} & Theory & Experiment \\  &
\multicolumn{2}{c}{} \\ \hline & & \\ Au-Au at RHIC at
$\sqrt{s_{NN}}=200$ GeV & 0.99 \footnote{For the modified
definition of $E_{T}$, \emph{i.e.} $E_{i}+m_{N}$ is taken instead
of $E_{i}$ for antibaryons, see Eq.~(\protect\ref{Etdef}).} &
$0.87 \pm 0.06$ \protect\cite{Bazilevsky:2002fz} \\ & & $1.03 \pm
0.08$ \footnote{Author calculations with the use of experimental
values given in Table\,\protect\ref{Table1} and the denominator
expressed as the sum of integrated charged hadron yields.} \\
Au-Au at RHIC at $\sqrt{s_{NN}}=130$ GeV & 0.91 & $0.81 \pm 0.06$
\protect\cite{Adcox:2001ry} \\ & & $0.89 \pm 0.09\;^{b}$ \\ Pb-Pb
at SPS
& 0.94 & $0.78 \pm 0.21$ \protect\cite{Aggarwal:2000bc} \\ & & \\
Au-Au
at AGS & 0.83 & $0.72 \pm 0.08$ \protect\cite{Barrette:1994kr} \\
& & \\ Si-Pb at AGS & 0.63 & 0.52-0.54
\protect\cite{Barrette:1994kr} \\ & & \\ \hline
\end{tabular}
\end{ruledtabular}
\end{table}
\begin{figure}
\includegraphics{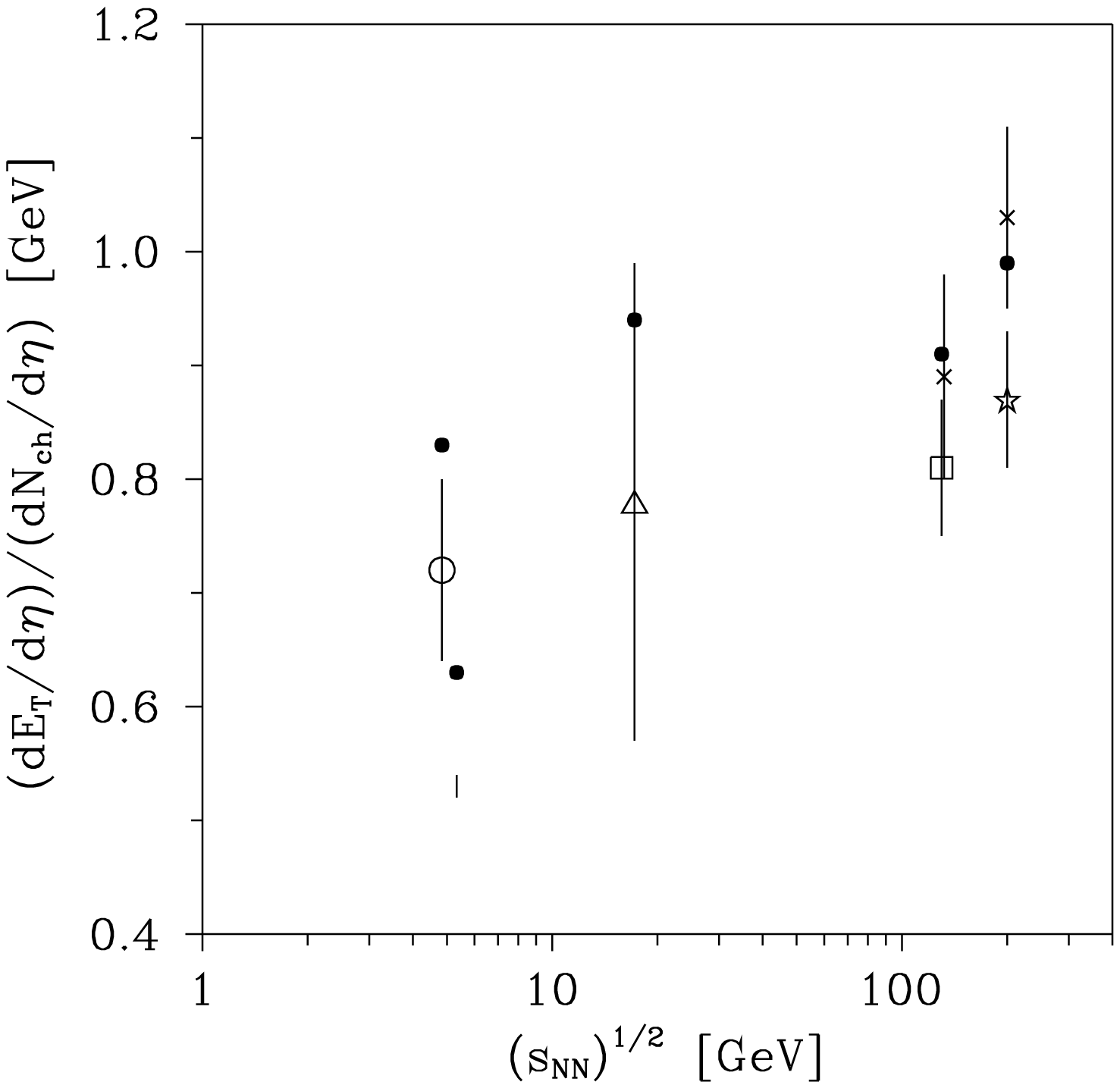}
\caption{\label{Fig.1.} Values of the transverse energy per
charged particle at midrapidity for the most central collisions.
Black dots denote evaluations of the ratio in the framework of the
present model (the second column of Table\,\protect\ref{Table2}).
Also data points for AGS \protect\cite{Barrette:1994kr} (a circle
for Au-Au and a vertical bar for Si-Pb), SPS
\protect\cite{Aggarwal:2000bc} (triangle), RHIC at
$\sqrt{s_{NN}}=130$ GeV \protect\cite{Adcox:2001ry} (square) and
RHIC at $\sqrt{s_{NN}}=200$ GeV \protect\cite{Bazilevsky:2002fz}
(star) are depicted. For RHIC, points with the sum of integrated
charged hadron yields substituted for the denominator are also
depicted (crosses).}
\end{figure}
These results have been also depicted together with the data in
Fig.\,\ref{Fig.1.}. One can see that the relative positions of
theoretical points agree very well with the data, they are shifted
up only and this is the effect of the overestimation discussed
earlier.

\section {Conclusions}
\label{Conclud}

The expanding thermal hadron gas model has been used to reproduce
transverse energy and charged particle multiplicity pseudorapidity
densities and their ratio measured at AGS, SPS and RHIC. The
importance of the present analysis originates from the fact that
the transverse energy and the charged particle multiplicity are
\emph{independent observables}, so they can be used as new tools
to verify the consistency of predictions of a statistical model
for all colliders simultaneously. The predictions have been made
at the previous estimates of thermal and geometric freeze-out
parameters obtained from analyses of measured particle ratios and
$p_{T}$ spectra at AGS \cite{Braun-Munzinger:1994xr,Stachel:wh},
SPS \cite{Michalec:2001um,Broniowski:2002am} and RHIC
\cite{Baran:2003nm,Broniowski:2002nf}. The overall good agreement,
not only of the ratio but also absolute values of
$dE_{T}/d\eta\mid_{mid}$ and $dN_{ch}/d\eta\vert_{mid}$, with the
data has been achieved. And the observed discrepancies can be
explained reasonably. This strongly supports the idea that the
thermal expanding source is responsible for the soft part of the
particle production in heavy-ion collisions. In addition, the
description of various observables is consistent within one
statistical model.

There are more arguments in favour of the above statement. One
could think that this analysis is a kind of an internal
consistency check of various measurements. And such a check could
be done even in an model-independent way simply by integrating
spectra of stable particles. But it can not be done without any
external input. First, transverse momentum spectra are measured in
\emph{limited ranges}, so very important low-$p_{T}$ regions are
not covered by the data. Therefore, to obtain integrated yields
some extrapolations below and above the measured ranges are used.
In fact these extrapolations are only analytical fits without any
physical reasoning, but contributions from regions covered by them
account for $25 \%-40 \%$ of the yield \cite{Adcox:2001mf}. On the
other hand, a calorimeter acts very effectively in the low-$p_{T}$
range, namely pions with $p_{T} \leq 0.35$ GeV/c, kaons with
$p_{T} \leq 0.64$ GeV/c and protons and antiprotons with $p_{T}
\leq 0.94$ GeV/c are all captured \cite{Adcox:2001ry}. Since the
very accurate predictions for the transverse energy density at
midrapidity have been obtained (see Table\,\ref{Table1}), the
present analysis can be understood as an undirect proof that in
these unmeasurable $p_{T}$ regions spectra are also explicable by
means of the thermal source with flow and decays.

Moreover, it is impossible to check the consistency of the
transverse energy data because not all stable hadron spectra are
measured. This mainly concerns neutrons and $K_{L}^{0}$. Also it
is impossible to extract hadron decay photons from the photon
data. And again, the very good agreement of model estimates of the
transverse energy density at midrapidity with the data can be
interpreted as the strong argument that the production of neutral
stable particles can be described in terms of the expanding
thermal source with superimposed decays.

And as the last remark: in opposite to the transverse energy,
there is some inconsistency (of the order of $10 \%$) of the
independent measurements of the charged particle multiplicity with
the corresponding sums of integrated charged particle yields at
RHIC (see sect.~\ref{Finl}). But at the present stage of
investigation it is difficult to judge whether this inconsistency
has the physical or experimental reason.

\begin{acknowledgments}
The author gratefully acknowledges very stimulating discussions
with Wojciech Broniowski and Wojciech Florkowski. This work was
supported in part by the Polish Committee for Scientific Research
under Contract No. KBN 2 P03B 069 25.
\end{acknowledgments}

\end{document}